\documentclass[%
reprint,
superscriptaddress,
longbibliography,
preprintnumbers,
amsmath,amssymb,
aps,
prl,
floatfix,
]{revtex4-2}
\usepackage{xparse}
\usepackage{siunitx}

\usepackage[utf8]{inputenc}
\usepackage[T1]{fontenc}
\usepackage[english]{babel}
\addto\captionsenglish{}
\usepackage{bm}

\usepackage{braket}

\usepackage[mode=buildnew]{standalone}

\usepackage[maxfloats=256]{morefloats}
\usepackage[table]{xcolor}
\definecolor{elsa_green}{HTML}{B4C79D}
\definecolor{elsa_green_2}{HTML}{C2D1B2}

\usepackage{hyperref}
\begin{document}

\preprint{}



\title{Gate-tunable single terahertz meta-atom ultrastrong light-matter coupling}

\date{\today}

\author{Elsa Jöchl}
\author{Anna-Lydia Vieli}
\author{Lucy Hale}
\author{Felix Helmrich}
\affiliation{Institute of Quantum Electronics, ETH Zürich, Zürich 8093, Switzerland}
\author{Deniz Turan}
\author{Mona Jarrahi}
\affiliation{Department of Electrical $\&$ Computer Engineering, University of California Los Angeles (UCLA), Los Angeles, CA, USA }
\author{Mattias Beck} 
\author{Jérôme Faist}
\author{Giacomo Scalari}
\affiliation{Institute of Quantum Electronics, ETH Zürich, Zürich 8093, Switzerland}
\begin{abstract}
We study the electrical tunability of ultrastrong light-matter interactions between a single terahertz circuit-based complementary split ring resonator (cSRR) and a two-dimensional electron gas. For this purpose, transmission spectroscopy measurements are performed under the influence of a strong magnetic field at different set points for the electric gate bias. The resulting Landau polariton dispersion depends on the applied electric bias, as the gating technique confines the electrons in-plane down to extremely sub-wavelength dimensions as small as $d = \qty{410}{nm}$. This confinement allows for the excitation of standing plasma waves at zero magnetic field and an effective tunability of the electron number coupled to the THz resonator. This allows the normalized coupling strength to be tuned in-situ from $\eta = 0.46$ down to $\eta = 0.18$. This is the first demonstration of terahertz far-field spectroscopy of an electrically tunable interaction between a single terahertz resonator and electrons in a GaAs quantum well heterostructure.

\end{abstract}

\maketitle


\section{Introduction}
\begin{figure*}
    \includegraphics[scale = 0.9]{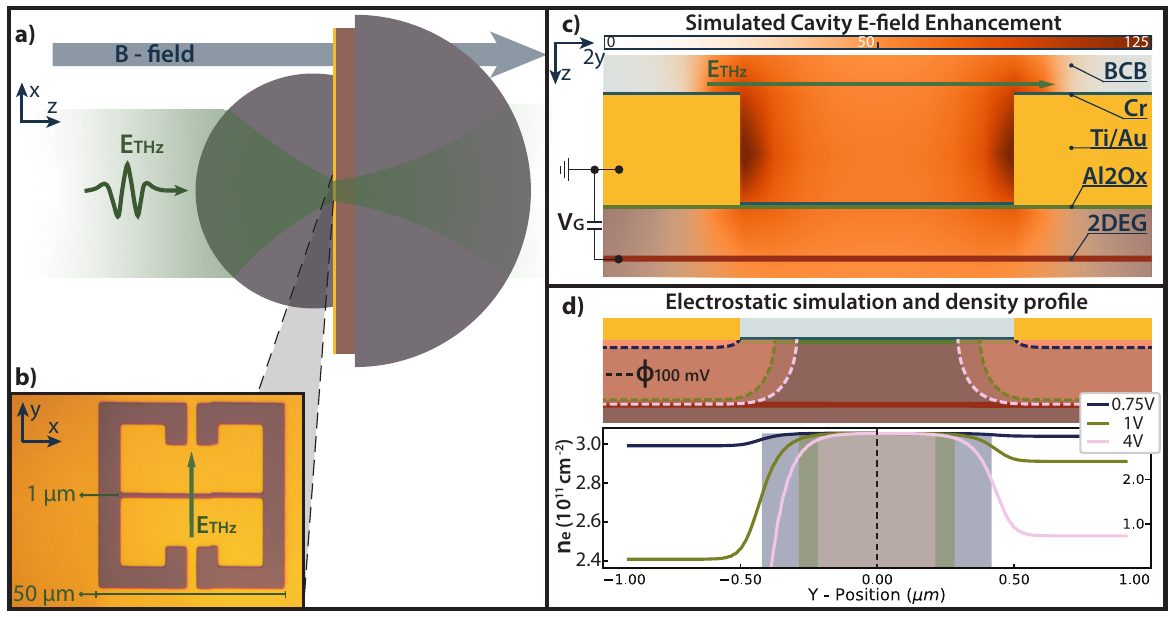}
    \caption{a) Schematic of the sample mounted with aSIL system. b) An optical image of the cSRR. c) Schematic cut view of the sample across the cSRR gap, overlaid with the electric field confinement at zero electric bias (simulated using FEM). d) Simulated DC field distribution (equipotential lines at $\Phi = 100 \,\textrm{mV}$ indicated with dotted lines) and lateral narrowing of the 2DEG strip due to an increase in DC bias. The aspect ratio between the x and y axis in the schematic panels of c) and d) is set to 1:2 to improve visibility of the thin layers. The computed change in the electron density for increasing potential is shown in the bottom of panel d), according to a Thomas-Fermi distribution. To show the asymptotic values for all cases, the y-axis is squeezed in the right side of the plot.}
    \label{fig:sample_schematic}
\end{figure*}

Ultrastrong light-matter interaction has been studied extensively in the terahertz (THz) frequency range \cite{kockum_ultrastrong_nodate, forn-diaz_ultrastrong_2019}. At these low energies, the Rabi-frequency of the light-matter coupled system can readily reach a significant fraction of the system's uncoupled eigenfrequencies, which is the defining condition for ultrastrong coupling. This non-perturbative hybridization of light and matter has several fundamental implications, for example, on the ground state of the system \cite{ciuti_quantum_2005}, which hosts virtual photons. Recent experimental and theoretical efforts have highlighted the potential of such ultrastrong light-matter interactions to modify material properties\cite{lu_cavity_2025, anappara_signatures_2009, appugliese_breakdown_2022, enkner_tunable_2025, kim_observation_2025, ashida_quantum_2020}. 

One way of achieving ultrastrong coupling is by confining the THz electric fields strongly in sub-wavelength modes through the use of circuit-based resonators, so-called split-ring resonators (SRRs) \cite{Pendry_IEEE_1999,scalari_ultrastrong_2013, bayer_terahertz_2017, mornhinweg_mode-multiplexing_2024, halbhuber_non-adiabatic_2020}. Furthermore, intra-band transitions in semiconductor heterostructures at these frequencies can be easily integrated with these SRR modes \cite{jeannin_ultrastrong_2019,todorov_ultrastrong_2010}. Here, we couple Landau level transitions in a GaAs single quantum well heterostructure to a single complementary SRR (cSRR) mode by applying a magnetic field perpendicular to a two-dimensional electron gas (2DEG) in a GaAs quantum well heterostructure. 

There has been extensive effort in studying ultrastrong coupling with few electrons in the THz regime \cite{keller_few-electron_2017, jeannin_ultrastrong_2019, kuroyama_coherent_2024, rajabali_ultrastrongly_2022}.
In this case, the electronic excitation can no longer be treated as a harmonic, quasi-bosonic mode, and its fermionic nature starts to play a role. The corresponding anharmonicity is predicted to result in a strikingly different polariton dispersion \cite{todorov_few-electron_2014, casanova_deep_2010}.

One possibility to  reduce the interaction to a small ensemble of electrons is by fabricating meta-atoms with a very low mode volume, as done in reference \cite{keller_few-electron_2017}. This approach, however, does not allow for in-situ tunability of the interaction. Another method is to electrically tune the electron system via a gating technique \cite{paravicini-bagliani_gate_2017}. In typical experiments measuring large ($n>100$) arrays of collectively coupled systems, this requires uniform gating in each individual subwavelength region across a large area, which is technologically challenging. Recently, we have demonstrated far-field measurements of a single cSRR by using an asymmetric solid immersion lens (aSIL) system mounted in direct contact with the sample \cite{rajabali_ultrastrongly_2022}. In this work, we combine this single meta-atom spectroscopy technique with in-situ electrical modification of the electron system below the single cSRR. 

To tune the matter system, a gate bias is applied between the 2DEG and the metallic cSRR plane. By applying such a spatially inhomogeneous gate bias, the 2DEG is mainly depleted in the regions below the resonator surface, which creates a confinement in the shape of the resonator openings. Increasing the gate bias, therefore, enhances the in-plane confinement of the electron system. As a result, the overlap factor between the light and matter systems is no longer governed by the optical mode volume, but rather by the depletion-induced confinement of the electron system. This allows for dynamical tuning of the light-matter coupling strength, which depends on the number of electrons: $\hbar \Omega_R = \vec{d}_{ij} \, \vec{\varepsilon}_{vac} \, \sqrt{N_{\text{e}}}$, with $N_{\text{e}}$ the number of optically active electrons, $\vec{d}_{ij}$ the transition dipole moment of the matter system, and $\vec{\varepsilon}_{vac}$ the vacuum electric field of the resonator mode. Furthermore, shaping the 2DEG into the specific shape of the cSRR makes it possible to couple to standing plasma waves which appear at zero magnetic field and depend on the gate bias.


\section{Fabrication and Measurement Methods}


The sample is processed on a GaAs/AlGaAs single triangular quantum well at \qty{90}{nm} depth. This structure is grown in-house using molecular beam epitaxy, and exhibits a nominal electron sheet density of $n_e = 3 \times 10^{11}\, \mathrm{cm^{-2}}$ at zero electric field bias. 

In order to electrically contact the quantum well, ohmic contacts are established using a standard annealing process. 
To ensure electrical insulation of the 2DEG and avoid leakage currents through the GaAs cap layer, a thin layer of alumina (Al\textsubscript{2}O\textsubscript{3})
is deposited using atomic layer deposition. 
The alumina is etched down to a thickness of $t \leq$ \qty{10}{nm} in the area where the cSRR will later be placed to ensure maximal overlap between the 2D electron system and the resonant cavity mode. The remaining non-zero thickness of $\mathrm{Al}_{2}\mathrm{O}_{x}$ ensures that the GaAs cap layer underneath the cSRR is protected in the later etching steps, leaving the underlying quantum well intact. Afterwards, the 
resonator plane is deposited using a lithographic lift-off process. It consists of a circle of diameter $d =$ \qty{4}{mm} with a single cSRR placed in the center, and contact pads placed next to the ohmic contacts mentioned above. The cSRR is designed to be resonant at a frequency of $f = \qty{270}{GHz}$. In an attempt to form a homogeneous gate, we deposited a 4 nm thin Cr layer (similarly to reference \cite{paravicini-bagliani_gate_2017}) to deplete the 2DEG density uniformly across the sample, according to a Schottky-contact enabled depletion mechanism. As we will show, the measurements nonetheless indicate a lateral 2DEG confinement, leading us to the conclusion that the thin Cr layer within the cSRR gap is electrically disconnected from the cSRR plane used for gating.

In the last step, a protective \qty{2}{\mu m} thick layer of benzocyclobutene polymer (BCB) is spin coated onto the sample. As first shown in our previous work \cite{rajabali_ultrastrongly_2022}, we can utilize an aSIL system, shown in Figure \ref{fig:sample_schematic}a, to excite a single meta-atom mode with far-field THz radiation, which will be coupling to our tunable 2DEG platform. The BCB-layer acts as a buffer to protect the sample from shearing damage when the lenses are put in place. This is necessary to ensure insulation between the electrically conductive layers. 

After mounting the sample together with the aSILs in a cleanroom environment, it is placed in a cryomagnet system, where we perform THz transmission time-domain spectroscopy (TDS) at \qty{3}{K} as a function of  the magnetic field with different gate biases. The utilised THz source is presented in reference \cite{turan_impact_2017}, and driven by a mode-locked Ti:Sapph oscillator at 800 nm. The THz field is detected by electro-optic sampling in ZnTe. The mounted aSIL structure along with the E-field polarization and direction of the magnetic field are indicated in Figure \ref{fig:sample_schematic}a.
An optical picture of the fabricated cSRR can be seen in Figure \ref{fig:sample_schematic}b, along with a schematic side cut view along the cSRR gap in Figure \ref{fig:sample_schematic}c, where the simulated electric field confinement is overlaid.

\begin{figure}

    \includegraphics[scale = 1.25]{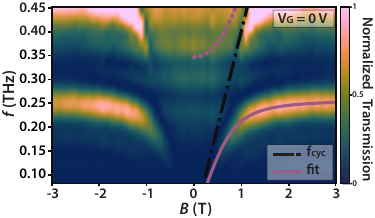}
    \caption{THz TDS measurements of the single cSRR sample shown in Figure \ref{fig:sample_schematic}, performed at $3\, \mathrm{K}$ without gate bias. The solid purple curve represents the fitted lower polariton branch according to the Hopfield model, while the dotted line represents the expected upper polariton dispersion. We observe broadened transmission instead of a localized branch due to plasmonic broadening. The black line shows the bare cyclotron dispersion corresponding to an effective mass of $m_{\text{eff}} = 0.07 \, m_{\text{e}}$.}
    \label{fig:results0V}
\end{figure}

\begin{figure*}
    \includegraphics[scale = 0.98]{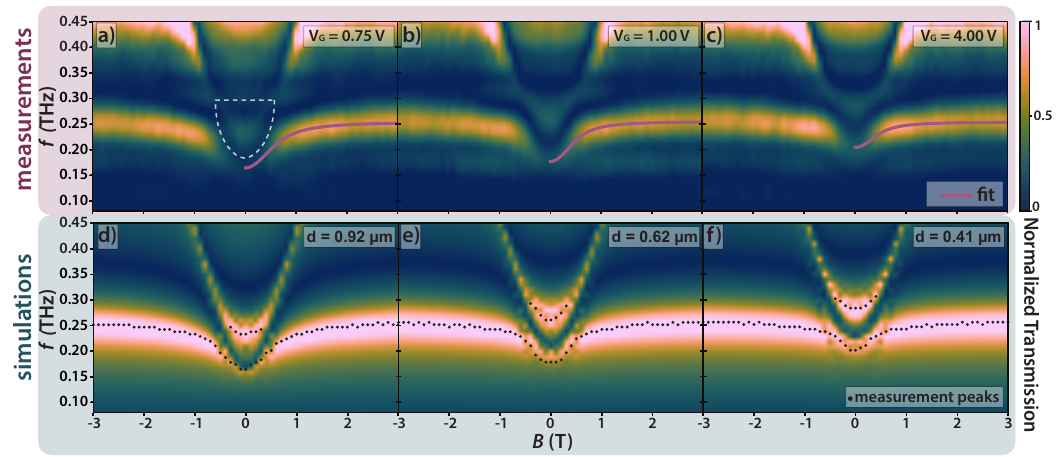}
    \caption{a)-c): THz TDS measurements of the single cSRR shown in Figure \ref{fig:sample_schematic}, performed at $3\, \mathrm{K}$ for varying back-gate biases. The purple curve represents the fitted lower polariton branch according to the Hopfield model (Eq.\,\eqref{eq:hopfield}). As discussed above, we do not expect to observe an upper polariton branch due to plasmonic broadening, hence why the UP fit curves are omitted. d)-f): Simulations of the sample with varying depletion lengths, overlaid with the transmission peaks extracted from the measured data for comparison.}
    \label{fig:results1}
\end{figure*}

\section{Results}
The measured spectrum for zero electric bias is shown in Figure \ref{fig:results0V}. The transmission map shows the polariton dispersion as a function of magnetic field, spanning from \qty{-3}{T} to \qty{3}{T}. The spectral map is extended to negative magnetic fields to better visualize the features appearing at $B = \qty{0}{T}$. The lower polariton (LP) branches are symmetric in $B$, well defined, and bend from \qty{150}{GHz} up to the cavity frequency $f = \qty{255}{GHz}$ around the anticrossing at $|B| = \qty{0.7}{T}$. The upper polariton (UP) appears as a broadband transmission region between \qty{300}{GHz} and \qty{450}{GHz} at magnetic fields below $|B| = \qty{1}{T}$. The transmission peak appearing below the slope of the cyclotron resonance upwards of $f = \qty{400}{GHz}$ is an artifact of the normalization procedure, and is not of interest for the further analysis.

In order to understand the measured spectra, we can compare the measured data to the expected Hopfield-like dispersion for a conventional ultrastrong coupling interaction between Landau-quantized electrons and a cavity mode. We expect two polariton branches with frequencies given by
\cite{hagenmuller_ultrastrong_2010}:
\begin{equation} \label{eq:hopfield}
    \omega_{\text{LP}}^{\text{UP}} = \sqrt{\frac{1}{2} (w_{\text{cyc}}^2 + 4 \Omega_{\text{R}}^2 + \omega_{\text{cav}} ^2 \pm G)} 
\end{equation}
with the polariton gap $G$
\begin{equation}
    G = \sqrt{-4 \omega_{\text{cyc}}^2 \omega_{\text{cav}}^2 + (-\omega_{\text{cyc}}^2 - 4\Omega_{\text{R}}^2-\omega_{\text{cav}}^2)}.
\end{equation}
Here, $\omega_\text{cyc}$ denotes the bare cyclotron transition, $\omega_{\text{cav}}$ the cSRR mode, and $\Omega_R$ the vacuum Rabi frequency.

While the LP branches follow this expected dispersion behavior, the UP is affected by polaritonic nonlocality and does not exhibit a single well-defined frequency \cite{rajabali_polaritonic_2021, endo_cavity-mediated_2025}. The dispersion of the UP broadens because the electrons couple to a continuum of plasma excitations with different in-plane momenta made accessible via the fundamental cSRR mode confined in the gap of size $W = \qty{1}{\mu m}$. As the gap size decreases, the highest possible photonic momentum mode $k \propto \frac{1}{W}$ also increases. The upper bound for the continuum is given by the magnetoplasmon frequency
\begin{equation}
    \omega_{\text{MP}} = \sqrt{\frac{n_e \, e^2 \, \pi }{2m^*\epsilon_{\text{eff}}W}}
    \label{eq:MP}
\end{equation}
with $n_e$ the electron sheet density, $e$ the electron charge, $m^*$ the electron effective mass, and $\epsilon_{\text{eff}}$ the effective dielectric constant of the medium. If the UP appears below this frequency, it couples to the continuum and broadens. For a resonator with a gap of $1 \, \mathrm{\mu m}$ coupling to a 2DEG of sheet density $n_e = 3 \times10^{11}\mathrm{cm}^{-2} $, this magnetoplasmon continuum has an upper limit of $f = 650 \, \mathrm{GHz}$, well above the expected UP frequency. A UP with frequencies below this limit will therefore appear as smeared out, as is reflected in the measurement. 

The Hopfield model provides an expectation of the LP given by Eq.\, \ref{eq:hopfield}. By fitting the measured LP data to this curve, we can quantify the normalized coupling ratio to be $\eta = \Omega/\omega = 0.456$ with a cavity frequency of $255 \, \mathrm{GHz}$. 

After the calibration measurement at $V_G = 0$, we can study how an applied gate bias modifies the light-matter interaction. In Figure \ref{fig:results1}a-c we present measurements of the same sample at increasing gate biases. These measurements were performed in separate magnetic field sweeps. To account for small changes in the optical alignment, we performed calibration measurements at zero magnetic field and gate bias before each scan. Starting at a gate bias of $V_G =$ \qty{0.75}{V}, we observe a strong modification of the LP dispersion. Firstly, we observe that the LP branch blue-shifts in frequency below the anticrossing, with a non-zero asymptotic value at $B = \qty{0}{T}$. This asymptotic value furthermore blue shifts from \qty{180}{GHz} up to \qty{200}{GHz} as the voltage bias is increased to \qty{4}{V}. Secondly, a transmission maximum forms slightly above the LP asymptotic value, which peaks at $f = \qty{225}{GHz}$ at the lowest gate bias, and blueshifts with increasing bias up to $f = \qty{280}{GHz}$. This additional feature is indicated by a white dotted line in Figure \ref{fig:results1}a, and will furthermore be referred to as the M1 mode.

To gain insights about the coupling strength at varying gate biases, the Hopfield model can be used to fit the LP branches. However, in this case, the cyclotron frequency $\omega_{\text{cyc}}$ in Equation \ref{eq:hopfield} is renormalized by the magnetoplasmon frequency given in Equation \ref{eq:MP}, to account for the non-zero asymptote of the LP branches at $B = \qty{0}{T}$, as done in reference \cite{paravicini-bagliani_gate_2017}: $\tilde{\omega}_{\text{cyc}} = \sqrt{\omega_{\text{cyc}}^2+\omega_{\text{MP}}^2}$. 
The normalized coupling ratios estimated with this method are decreasing with increasing gate bias $\eta_{\text{0.75V}} = 0.333$, $\eta_{\text{1V}} = 0.231$, and $\eta_{\text{4V}} = 0.184$, with the fitted magnetoplasmon frequencies $f_{\text{0.75V}} = 218\, \mathrm{GHz}$, $f_{\text{1V}} = 210\, \mathrm{GHz}$ and $f_{\text{4V}} = 240\, \mathrm{GHz}$.

The emerging M1 mode 
is not trivially explained. It cannot originate from the UP branch, since its center frequency lies below the cavity frequency at the lowest applied gate bias V$_G$ = \qty{0.75}{V}. Furthermore, the excitation blue-shifts with increasing bias. As the coupling strength decreases with increasing bias, we would expect a UP branch to red-shift. The M1 mode in fact stems from the way the gate bias is applied to the 2DEG. As the voltage bias is applied directly using the resonator plane, the cSRR imprints its shape onto the 2DEG, which confines the electron system laterally. 
As a result, the higher the applied DC bias, the more tightly the 2DEG will be confined in-plane. Finite element simulations performed in COMSOL support this intuitive picture. 

First, we can investigate the 2D electrostatic distribution of the electric potential along a sidecut of the cSRR gap with increasing applied potentials. The results of such a simulation are shown in Figure \ref{fig:sample_schematic}d, where the respective equipotential lines at \qty{100}{mV} are indicated, representing the cutoff of electric potential for different voltage biases. From this, we can calculate the electron density $n_e$ along the $z$-position of the 2DEG according to a Thomas-Fermi distribution \cite{beenakker_quantum_1991}. The density starts to decrease by 1\% at the position where the electric potential drops, which generates an effective electron channel of width $d$, confined beneath the cSRR gap. This channel varies in width from roughly \qty{90}{\mu m} down to \qty{40}{\mu m} at $V_G =$ \qty{4}{V}. The exact simulation parameters and potential distributions are detailed in the Supplementary Material. 

Then, the polariton dispersion is simulated with varying confinement strengths, corresponding to the obtained estimated effective widths $d$ of the electrostatic simulation. The simulation model accounts for a 2DEG (modelled as a gyrotropic medium) with an abrupt cutoff in the shape of the cSRR, placed underneath the actual metallic cSRR. The width of this cSRR-shaped 2DEG is varied to model the effective confinement width of the 2DEG in the central cSRR gap. The simulated magnetic field is then swept to tune the cyclotron dispersion and replicate the measured transmission maps as closely as possible. The optimal simulated spectral maps correspond to channel widths $d_{\text{0.75V}}$ = \qty{0.92}{\mu m}, $d_{\text{1V}}$ = \qty{0.62}{\mu m}, and $d_{\text{4V}}$ = \qty{0.41}{\mu m}, and are shown in Figure \ref{fig:results1}c-d. These spectra show good agreement with the measurements as well as the expected widths from the electrostatic simulations for the corresponding bias voltages. The transmission peaks extracted from the measurements are overlaid with the simulated data for direct comparison. We can therefore conclude that the emerging spectral feature stems from the inhomogeneous confinement of the 2DEG.



\begin{figure}
    \centering
    \includegraphics[scale=0.44]{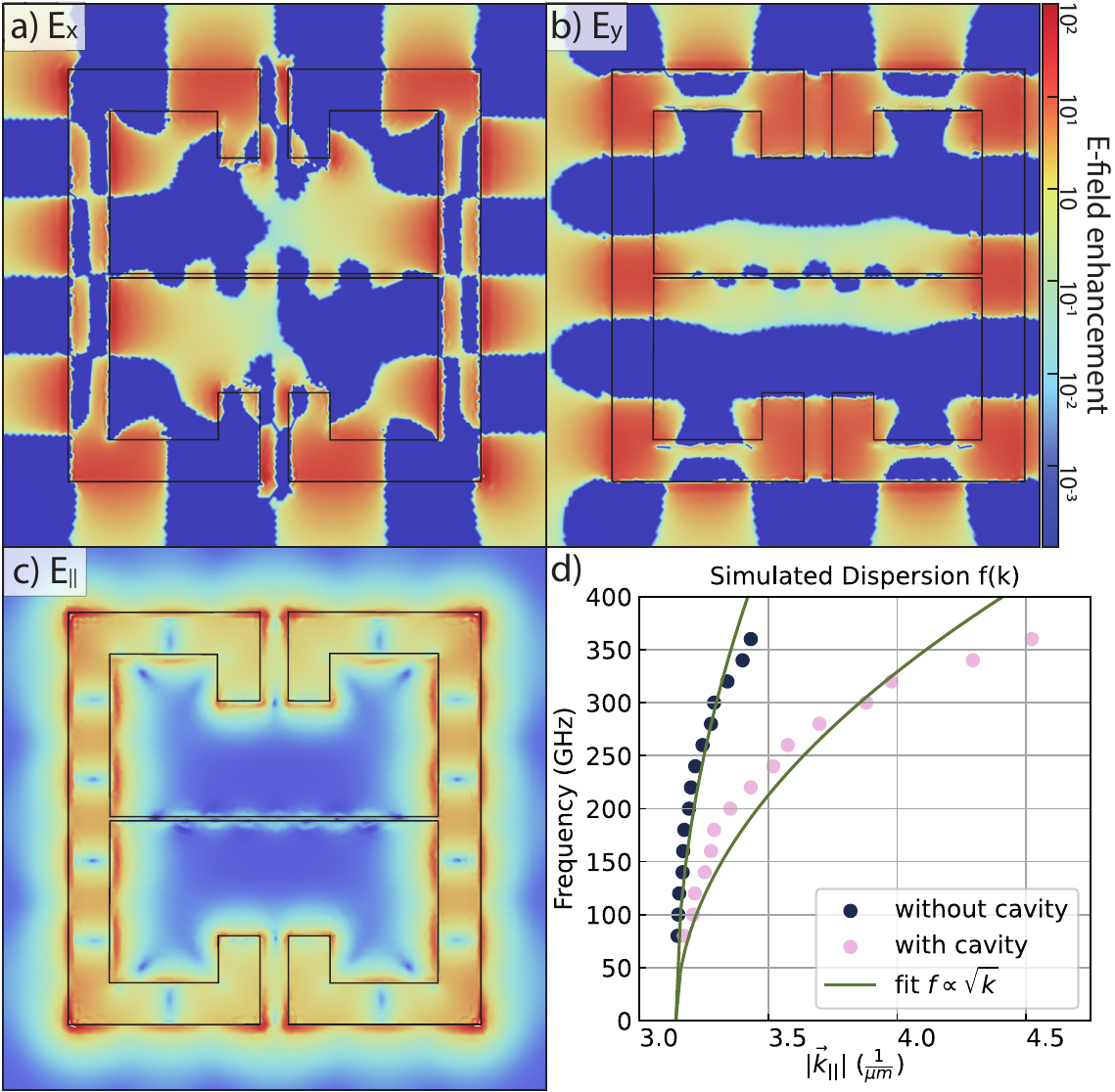}
    \caption{Plasma excitations of the bare 2DEG. a-c) Normalized $x$-,$y$- and in-plane component of the simulated electric field for a 2DEG confined to the cSRR geometry without the resonator plane. The cutoff of the cSRR-shaped 2DEG is depicted with a black line. Represented at \qty{200}{GHz} and a confinement width of d = \qty{0.41}{\mu m} in the central gap. d) The resulting dependence of the frequency on the in-plane wavevector, as inferred from the field distribution. Overlaid with a square-root fit for comparison.}
    \label{fig:fig4}
\end{figure}

By constraining the 2DEG spatially in this way, standing plasma wave excitations are induced within the 2DEG, which hybridize with the LP branch. An example simulation of such a standing wave pattern is shown in Figure \ref{fig:fig4}a-c. In this case, only the laterally confined 2DEG is simulated, and the cSRR is omitted to investigate the plasma excitations irrespective of their hybridization with a cavity mode. 

The $E_x$ and $E_y$ components of the in-plane electric field (normalized to the input electric field) are plotted in Figure \ref{fig:fig4}a and b respectively, and exhibit a clear standing wave pattern. The full in-plane electric field distribution is shown in Figure \ref{fig:fig4}c. The plasma excitations exhibit the same dependence on the wave vector as the magnetoplasmon described in Equation \ref{eq:MP}, namely $\omega \propto \sqrt{k}$, as established by estimating their wavelengths from the simulated in-plane electric field distributions at different frequencies. The resulting values are reported in Figure \ref{fig:fig4}d. 

Finally, we can estimate the number of ultrastrongly coupled electrons and compare the finite-bias cases with the nominal measurement at \qty{0}{V} bias. At zero bias, the number of interacting electrons is given by \cite{keller_few-electron_2017}:
\begin{equation}
    N_{\text{e}}^0 = \frac{e B}{h} \times S_{\text{eff}} = 7860
    \label{eq:Ne}
\end{equation}
where $S_{\text{eff}} = 41 \,\mathrm{\mu m^2}$ is the effective cavity surface at the position of the electron system, computed analytically with finite element simulations. The magnetic field $B =$\qty{0.8}{T} is the field where the cyclotron mode anticrosses with the cavity mode.

Equation \ref{eq:Ne} for the electron number becomes inaccurate when a magnetoplasmon continuum forms, as the Landau Level quantization of the electrons is now modified by non-local effects \cite{keller_few-electron_2017}. However, we can relate the Rabi frequency to the number of electrons $\Omega_{\text{R}} \propto \sqrt{N_e}$, and take the ratio of the coupling strength at different gate bias setting. This procedure yields a minimal electron number of $N_{\text{e}}  = (\frac{\eta_{4V}}{\eta_{0}})^2 \times N_\text{e}^0= 1260$ at \qty{4}{V} gate bias. Exact numbers for all measurements are given in Table \ref{tab:overlap}.


To further verify the decrease in the fitted coupling strengths and the subsequently obtained decrease in the electron number $N_\text{e}$, the light-matter coupling can be studied in terms of the overlap factor

\begin{equation}
    \Gamma_V = \frac{1}{\Gamma_0} \frac{\int_{S_V} \big\vert E_{xy}\big\vert \,dS}{\int_{S}  \big\vert E_{xy}\big\vert \,dS\, }.
\end{equation}

It is defined as the ratio of the absolute in-plane electric field $\vert E_{xy}\vert $ within the surface of the confined 2DEG channel $$S_V = \int\,dx\int_{-d/2}^{d/2} dy \, \vert E_{xy}\vert \, \bigg\vert_{z=z2DEG}$$ normalized to the entire electric field in the 2DEG plane (over the surface $S$ extending beyond the 2DEG constraints). The normalization factor $\Gamma_0$ is chosen so that the overlap factor is equal to unity in the case of no electron confinement. Finite element simulations provide the electric field values to compute the overlap factor for the gate biases applied in the measurements. The resulting overlap factors are reported in Table \ref{tab:overlap}. Since the coupling strength depends on the magnitude of the electric field, the number of ultrastrongly coupled electrons scales in the same fashion as $\Gamma_V$, which shows good agreement with the analysis performed via the coupling strengths, and yields a minimal reached electron number of $N^*_\text{e} = 1440$.

\begin{table}
\begin{tabular}{|c | c |c |c |c| c |c|} 
\hline
\rowcolor{elsa_green} 
Bias & Width d & $\eta$ & $\big(\frac{\eta}{\eta_0}\big)^2$ & $\Gamma_V$ & $N_\text{e}$ & $N_\text{e}^*$ \\[1.5ex] 
\rowcolor{elsa_green_2} V & $\mu m$ & & & & $N_\text{e}^0 \times \big(\frac{\eta}{\eta_0}\big)^2$ & $N_0 \times \Gamma_V$\\ [1.5ex] 
\hline
0 & - & 0.456 & 1 & 1 & 7860 & 7860\\ 
0.75 & 0.92 & 0.333 & 0.52 & 0.55 & 4090 & 4330\\ 
1.00 & 0.62 & 0.231 & 0.26 & 0.31 & 2040 & 2400\\ 
4.00 & 0.41 & 0.184 & 0.16 & 0.18 & 1260 & 1440\\ 
\hline
\end{tabular} 
\caption{Summary of the applied gate biases, simulated electron channel widths, coupling strengths obtained from Hopfield fits, the coupling strength relative to the zero gate bias measurement, the electron number $N_{\text{e}}$ computed via those ratios, the overlap factors, and electron numbers obtained via the overlap factors $N_{\text{e}}^*$.} 
\label{tab:overlap} 
\end{table} 

Electrical gating, therefore, provides a method to tune the number of ultrastrongly coupled electrons by almost an order of magnitude. The limitation is currently given by leakage currents breaking through the insulating layer of alumina. Spatially confining the electron system through etching, and combining such a structure with a cSRR with a lower mode volume as done in \cite{keller_few-electron_2017}, would solve these limitations. In such a resonator, the electron number could be decreased down to $N_{\text{e}} < 10$ by gate-tuning the light-matter coupling strength to $\eta = 0.1$ with the presented method.
\section{Conclusion}
We have successfully performed in-situ modification of the ultrastrong light-matter interaction between a single THz meta atom mode and a 2DEG, tuning the electron number by almost an order of magnitude. We have shown that spectroscopy on these tunable systems is possible by way of an aSIL system combined with an optimized fabrication process. The observed polariton dispersion is strongly modified by the application of an inhomogeneous voltage bias, which induces standing plasma waves in the two-dimensional electron system. These standing waves have been examined at different confinement widths and frequencies, and show a dispersion relation similar to conventional plasma excitations in 2D electron systems. The number of ultrastrongly coupled electrons and the coupling strength decrease as a function of increasing gate bias, which is corroborated by the study of the overlap between the cavity mode and the confined electronic mode.

This experiment constitutes a stepping stone for further experiments utilizing single-resonator THz spectroscopy with gate-tunable devices. Preemptively shaping underlying semiconductor-based electron systems into Hall Bars would assist in characterizing the electron density via electronic transport. This can further facilitate the application of a homogeneous gate bias to uniformly deplete the electron system. Being able to modify matter systems via an electrical gate at deeply sub-wavelength dimensions opens up the possibility to perform ultrastrong coupling experiments in a variety of unexplored platforms such as van der Waals heterostructures, as has recently been shown for bilayer graphene \cite{helmrich_cavity-driven_2025}. There is particular interest in monolayer graphene in a magnetic field, ultrastrongly coupled to a THz cavity mode. In this system, the emergence of a superradiant phase transition has been a long-standing subject of debate \cite{hagenmuller_cavity_2012, chirolli_drude_2012}. 

\section{Acknowledgments}
E.J. thanks Lorenzo Graziotto for fruitful discussions. E.J and F.H. thank Shima Rajabali for her contributions in the development of the aSIL measurement technique. We acknowledge the cleanroom facility FIRST at ETH Zurich. 

\section{Funding}
E.J., L.H., M.B., J.F., and G.S. acknowledge funding by the Swiss National Science Foundation (SNF) (Grant number 10000397).
F.H. acknowledges support from the Swiss National Science Foundation (SNF) (Grant number  200020 207520). D.T. and M.J. acknowledge funding by the US Department of Energy (Grant number DE-SC0016925).
\bibliography{references}

@article{jeannin_ultrastrong_2019,
	title = {Ultrastrong {Light}–{Matter} {Coupling} in {Deeply} {Subwavelength} {THz} {LC} {Resonators}},
	volume = {6},
	url = {https://doi.org/10.1021/acsphotonics.8b01778},
	doi = {10.1021/acsphotonics.8b01778},
	number = {5},
	urldate = {2025-09-09},
	journal = {ACS Photonics},
	author = {Jeannin, Mathieu and Mariotti Nesurini, Giacomo and Suffit, Stéphan and Gacemi, Djamal and Vasanelli, Angela and Li, Lianhe and Davies, Alexander Giles and Linfield, Edmund and Sirtori, Carlo and Todorov, Yanko},
	month = may,
	year = {2019},
	note = {Publisher: American Chemical Society},
	pages = {1207--1215},
}

@article{mornhinweg_mode-multiplexing_2024,
	title = {Mode-multiplexing deep-strong light-matter coupling},
	volume = {15},
	copyright = {2024 The Author(s)},
	issn = {2041-1723},
	url = {https://www.nature.com/articles/s41467-024-46038-9},
	doi = {10.1038/s41467-024-46038-9},
	language = {english},
	number = {1},
	urldate = {2025-07-28},
	journal = {Nature Communications},
	author = {Mornhinweg, Joshua and Diebel, Laura Katharina and Halbhuber, Maike and Prager, Michael and Riepl, Josef and Inzenhofer, Tobias and Bougeard, Dominique and Huber, Rupert and Lange, Christoph},
	month = feb,
	year = {2024},
	pages = {1847},
}

@misc{helmrich_cavity-driven_2025,
	title = {Cavity-{Driven} {Attractive} {Interactions} in {Quantum} {Materials}},
	url = {http://arxiv.org/abs/2408.00189},
	doi = {10.48550/arXiv.2408.00189},
	urldate = {2025-07-09},
	publisher = {arXiv},
	author = {Helmrich, F. and Adlong, H. S. and Khanonkin, I. and Kroner, M. and Scalari, G. and Faist, J. and Imamoglu, A. and Nova, T. F.},
	month = apr,
	year = {2025},
	note = {arXiv:2408.00189 [cond-mat]},
}

@article{hagenmuller_cavity_2012,
	title = {Cavity {QED} of the {Graphene} {Cyclotron} {Transition}},
	volume = {109},
	url = {https://link.aps.org/doi/10.1103/PhysRevLett.109.267403},
	doi = {10.1103/PhysRevLett.109.267403},
	number = {26},
	urldate = {2025-04-07},
	journal = {Physical Review Letters},
	author = {Hagenmüller, David and Ciuti, Cristiano},
	month = dec,
	year = {2012},
	note = {Publisher: American Physical Society},
	pages = {267403},
}

@article{chirolli_drude_2012,
	title = {Drude {Weight}, {Cyclotron} {Resonance}, and the {Dicke} {Model} of {Graphene} {Cavity} {QED}},
	volume = {109},
	url = {https://link.aps.org/doi/10.1103/PhysRevLett.109.267404},
	doi = {10.1103/PhysRevLett.109.267404},
	number = {26},
	urldate = {2025-04-07},
	journal = {Physical Review Letters},
	author = {Chirolli, Luca and Polini, Marco and Giovannetti, Vittorio and MacDonald, Allan H.},
	month = dec,
	year = {2012},
	note = {Publisher: American Physical Society},
	pages = {267404},
}

@article{rajabali_polaritonic_2021,
	title = {Polaritonic nonlocality in light–matter interaction},
	volume = {15},
	copyright = {2021 The Author(s), under exclusive licence to Springer Nature Limited},
	issn = {1749-4893},
	url = {https://www.nature.com/articles/s41566-021-00854-3},
	doi = {10.1038/s41566-021-00854-3},
	language = {english},
	number = {9},
	urldate = {2024-12-10},
	journal = {Nature Photonics},
	author = {Rajabali, Shima and Cortese, Erika and Beck, Mattias and De Liberato, Simone and Faist, Jérôme and Scalari, Giacomo},
	month = sep,
	year = {2021},
	note = {Publisher: Nature Publishing Group},
	pages = {690--695},
}

@article{hagenmuller_ultrastrong_2010,
	title = {Ultrastrong coupling between a cavity resonator and the cyclotron transition of a two-dimensional electron gas in the case of an integer filling factor},
	volume = {81},
	url = {https://link.aps.org/doi/10.1103/PhysRevB.81.235303},
	doi = {10.1103/PhysRevB.81.235303},
	number = {23},
	urldate = {2024-10-07},
	journal = {Physical Review B},
	author = {Hagenmüller, David and De Liberato, Simone and Ciuti, Cristiano},
	month = jun,
	year = {2010},
	note = {Publisher: American Physical Society},
	pages = {235303},
}

@article{paravicini-bagliani_gate_2017,
	title = {Gate and magnetic field tunable ultrastrong coupling between a magnetoplasmon and the optical mode of an {LC} cavity},
	volume = {95},
	url = {https://link.aps.org/doi/10.1103/PhysRevB.95.205304},
	doi = {10.1103/PhysRevB.95.205304},
	number = {20},
	urldate = {2024-09-25},
	journal = {Physical Review B},
	author = {Paravicini-Bagliani, Gian L. and Scalari, Giacomo and Valmorra, Federico and Keller, Janine and Maissen, Curdin and Beck, Mattias and Faist, Jérôme},
	month = may,
	year = {2017},
	note = {Publisher: American Physical Society},
	pages = {205304},
}

@article{kuroyama_coherent_2024,
	title = {Coherent {Interaction} of a {Few}-{Electron} {Quantum} {Dot} with a {Terahertz} {Optical} {Resonator}},
	volume = {132},
	url = {https://link.aps.org/doi/10.1103/PhysRevLett.132.066901},
	doi = {10.1103/PhysRevLett.132.066901},
	number = {6},
	urldate = {2024-04-04},
	journal = {Physical Review Letters},
	author = {Kuroyama, Kazuyuki and Kwoen, Jinkwan and Arakawa, Yasuhiko and Hirakawa, Kazuhiko},
	month = feb,
	year = {2024},
	note = {Publisher: American Physical Society},
	pages = {066901},
}

@article{appugliese_breakdown_2022,
	title = {Breakdown of topological protection by cavity vacuum fields in the integer quantum {Hall} effect},
	volume = {375},
	url = {https://www.science.org/doi/10.1126/science.abl5818},
	doi = {10.1126/science.abl5818},
	number = {6584},
	urldate = {2024-01-11},
	journal = {Science},
	author = {Appugliese, Felice and Enkner, Josefine and Paravicini-Bagliani, Gian Lorenzo and Beck, Mattias and Reichl, Christian and Wegscheider, Werner and Scalari, Giacomo and Ciuti, Cristiano and Faist, Jérôme},
	month = mar,
	year = {2022},
	note = {Publisher: American Association for the Advancement of Science},
	pages = {1030--1034},
}

@article{todorov_few-electron_2014,
	title = {Few-{Electron} {Ultrastrong} {Light}-{Matter} {Coupling} in a {Quantum} {LC} {Circuit}},
	volume = {4},
	url = {https://link.aps.org/doi/10.1103/PhysRevX.4.041031},
	doi = {10.1103/PhysRevX.4.041031},
	number = {4},
	urldate = {2024-01-11},
	journal = {Physical Review X},
	author = {Todorov, Yanko and Sirtori, Carlo},
	month = nov,
	year = {2014},
	note = {Publisher: American Physical Society},
	pages = {041031},
}

@article{ashida_quantum_2020,
	title = {Quantum {Electrodynamic} {Control} of {Matter}: {Cavity}-{Enhanced} {Ferroelectric} {Phase} {Transition}},
	volume = {10},
	shorttitle = {Quantum {Electrodynamic} {Control} of {Matter}},
	url = {https://link.aps.org/doi/10.1103/PhysRevX.10.041027},
	doi = {10.1103/PhysRevX.10.041027},
	number = {4},
	urldate = {2023-01-25},
	journal = {Physical Review X},
	author = {Ashida, Yuto and İmamoğlu, Ataç and Faist, Jérôme and Jaksch, Dieter and Cavalleri, Andrea and Demler, Eugene},
	month = nov,
	year = {2020},
	note = {Publisher: American Physical Society},
	pages = {041027},
}

@article{bayer_terahertz_2017,
	title = {Terahertz {Light}–{Matter} {Interaction} beyond {Unity} {Coupling} {Strength}},
	volume = {17},
	issn = {1530-6984},
	url = {https://doi.org/10.1021/acs.nanolett.7b03103},
	doi = {10.1021/acs.nanolett.7b03103},
	number = {10},
	urldate = {2023-01-25},
	journal = {Nano Letters},
	author = {Bayer, Andreas and Pozimski, Marcel and Schambeck, Simon and Schuh, Dieter and Huber, Rupert and Bougeard, Dominique and Lange, Christoph},
	month = oct,
	year = {2017},
	note = {Publisher: American Chemical Society},
	pages = {6340--6344},
}

@article{halbhuber_non-adiabatic_2020,
	title = {Non-adiabatic stripping of a cavity field from electrons in the deep-strong coupling regime},
	volume = {14},
	copyright = {2020 The Author(s), under exclusive licence to Springer Nature Limited},
	issn = {1749-4893},
	url = {https://www.nature.com/articles/s41566-020-0673-2},
	doi = {10.1038/s41566-020-0673-2},
	language = {english},
	number = {11},
	urldate = {2023-01-25},
	journal = {Nature Photonics},
	author = {Halbhuber, M. and Mornhinweg, J. and Zeller, V. and Ciuti, C. and Bougeard, D. and Huber, R. and Lange, C.},
	month = nov,
	year = {2020},
	note = {Number: 11
Publisher: Nature Publishing Group},
	pages = {675--679},
}

@article{keller_few-electron_2017,
	title = {Few-{Electron} {Ultrastrong} {Light}-{Matter} {Coupling} at 300 {GHz} with {Nanogap} {Hybrid} {LC} {Microcavities}},
	volume = {17},
	issn = {1530-6984},
	url = {https://doi.org/10.1021/acs.nanolett.7b03228},
	doi = {10.1021/acs.nanolett.7b03228},
	number = {12},
	urldate = {2023-01-12},
	journal = {Nano Letters},
	author = {Keller, Janine and Scalari, Giacomo and Cibella, Sara and Maissen, Curdin and Appugliese, Felice and Giovine, Ennio and Leoni, Roberto and Beck, Mattias and Faist, Jérôme},
	month = dec,
	year = {2017},
	note = {Publisher: American Chemical Society},
	pages = {7410--7415},
}

@article{anappara_signatures_2009,
	title = {Signatures of the ultrastrong light-matter coupling regime},
	volume = {79},
	issn = {0163-18291098-0121},
	url = {https://ui.adsabs.harvard.edu/abs/2009PhRvB..79t1303A},
	doi = {10.1103/PhysRevB.79.201303},
	urldate = {2022-11-14},
	journal = {Physical Review B},
	author = {Anappara, Aji A. and de Liberato, Simone and Tredicucci, Alessandro and Ciuti, Cristiano and Biasiol, Giorgio and Sorba, Lucia and Beltram, Fabio},
	month = may,
	year = {2009},
	note = {ADS Bibcode: 2009PhRvB..79t1303A},
	pages = {201303},
}

@article{ciuti_quantum_2005,
	title = {Quantum vacuum properties of the intersubband cavity polariton field},
	volume = {72},
	url = {https://link.aps.org/doi/10.1103/PhysRevB.72.115303},
	doi = {10.1103/PhysRevB.72.115303},
	number = {11},
	urldate = {2022-11-14},
	journal = {Physical Review B},
	author = {Ciuti, Cristiano and Bastard, Gérald and Carusotto, Iacopo},
	month = sep,
	year = {2005},
	note = {Publisher: American Physical Society},
	pages = {115303},
}

@article{scalari_ultrastrong_2013,
	title = {Ultrastrong light-matter coupling at terahertz frequencies with split ring resonators and inter-{Landau} level transitions},
	volume = {113},
	issn = {0021-8979},
	url = {https://aip.scitation.org/doi/full/10.1063/1.4795543},
	doi = {10.1063/1.4795543},
	number = {13},
	urldate = {2022-11-14},
	journal = {Journal of Applied Physics},
	author = {Scalari, G. and Maissen, C. and Hagenmüller, D. and De Liberato, S. and Ciuti, C. and Reichl, C. and Wegscheider, W. and Schuh, D. and Beck, M. and Faist, J.},
	month = apr,
	year = {2013},
	note = {Publisher: American Institute of Physics},
	pages = {136510},
}

@article{rajabali_ultrastrongly_2022,
	title = {An ultrastrongly coupled single terahertz meta-atom},
	volume = {13},
	copyright = {2022 The Author(s)},
	issn = {2041-1723},
	url = {https://www.nature.com/articles/s41467-022-29974-2},
	doi = {10.1038/s41467-022-29974-2},
	language = {english},
	number = {1},
	urldate = {2022-11-14},
	journal = {Nature Communications},
	author = {Rajabali, Shima and Markmann, Sergej and Jöchl, Elsa and Beck, Mattias and Lehner, Christian A. and Wegscheider, Werner and Faist, Jérôme and Scalari, Giacomo},
	month = may,
	year = {2022},
	note = {Number: 1
Publisher: Nature Publishing Group},
	pages = {2528},
}

@article{forn-diaz_ultrastrong_2019,
    title = {Ultrastrong coupling regimes of light-matter interaction},
    volume = {91},
    url = {https://link.aps.org/doi/10.1103/RevModPhys.91.025005},
    doi = {10.1103/RevModPhys.91.025005},
    number = {2},
    urldate = {2025-07-28},
    journal = {Reviews of Modern Physics},
    author = {Forn-Díaz, P. and Lamata, L. and Rico, E. and Kono, J. and Solano, E.},
    month = jun,
    year = {2019},
    pages = {025005},}

@ARTICLE{Pendry_IEEE_1999,
  author={Pendry, J.B. and Holden, A.J. and Robbins, D.J. and Stewart, W.J.},
  journal={IEEE Transactions on Microwave Theory and Techniques}, 
  title={Magnetism from conductors and enhanced nonlinear phenomena}, 
  year={1999},
  volume={47},
  number={11},
  pages={2075-2084},
  doi={10.1109/22.798002}}

@article{enkner_tunable_2025,
    title = {Tunable vacuum-field control of fractional and integer quantum {Hall} phases},
    volume = {641},
    copyright = {2025 The Author(s)},
    issn = {1476-4687},
    url = {https://www.nature.com/articles/s41586-025-08894-3},
    doi = {10.1038/s41586-025-08894-3},
    language = {english},
    number = {8064},
    urldate = {2025-07-28},
    journal = {Nature},
    author = {Enkner, Josefine and Graziotto, Lorenzo and Boriçi, Dalin and Appugliese, Felice and Reichl, Christian and Scalari, Giacomo and Regnault, Nicolas and Wegscheider, Werner and Ciuti, Cristiano and Faist, Jérôme},
    month = may,
    year = {2025},
    pages = {884--889},
}

@article{kim_observation_2025,
    title = {Observation of the magnonic {Dicke} superradiant phase transition},
    volume = {11},
    url = {https://www.science.org/doi/full/10.1126/sciadv.adt1691},
    doi = {10.1126/sciadv.adt1691},
    number = {14},
    urldate = {2025-09-29},
    journal = {Science Advances},
    author = {Kim, Dasom and Dasgupta, Sohail and Ma, Xiaoxuan and Park, Joong-Mok and Wei, Hao-Tian and Li, Xinwei and Luo, Liang and Doumani, Jacques and Yang, Wanting and Cheng, Di and Kim, Richard H. J. and Everitt, Henry O. and Kimura, Shojiro and Nojiri, Hiroyuki and Wang, Jigang and Cao, Shixun and Bamba, Motoaki and Hazzard, Kaden R. A. and Kono, Junichiro},
    month = apr,
    year = {2025},
    note = {Publisher: American Association for the Advancement of Science},
    pages = {eadt1691},
}

@article{todorov_ultrastrong_2010,
    title = {Ultrastrong {Light}-{Matter} {Coupling} {Regime} with {Polariton} {Dots}},
    volume = {105},
    doi = {10.1103/PhysRevLett.105.196402},
    number = {19},
    journal = {Physical Review Letters},
    author = {Todorov, Y.},
    year = {2010},
}

@misc{endo_cavity-mediated_2025,
    title = {Cavity-{Mediated} {Coupling} between {Local} and {Nonlocal} {Modes} in {Landau} {Polaritons}},
    url = {http://arxiv.org/abs/2509.05738},
    doi = {10.48550/arXiv.2509.05738},
    urldate = {2025-10-23},
    publisher = {arXiv},
    author = {Endo, Sae R. and Kim, Dasom and Liang, Shuang and Lee, Geon and Kim, Sunghwan and Covarrubias-Morales, Alan and Seo, Minah and Manfra, Michael J. and Lee, Dukhyung and Bamba, Motoaki and Kono, Junichiro},
    month = sep,
    year = {2025},
    note = {arXiv:2509.05738 [quant-ph]},
}

@article{turan_impact_2017,
    title = {Impact of the {Metal} {Adhesion} {Layer} on the {Radiation} {Power} of {Plasmonic} {Photoconductive} {Terahertz} {Sources}},
    volume = {38},
    issn = {1866-6906},
    url = {https://doi.org/10.1007/s10762-017-0431-9},
    doi = {10.1007/s10762-017-0431-9},
    language = {english},
    number = {12},
    urldate = {2025-10-31},
    journal = {Journal of Infrared, Millimeter, and Terahertz Waves},
    author = {Turan, Deniz and Corzo-Garcia, Sofia Carolina and Yardimci, Nezih Tolga and Castro-Camus, Enrique and Jarrahi, Mona},
    month = dec,
    year = {2017},
    pages = {1448--1456},
}

@incollection{beenakker_quantum_1991,
    series = {Semiconductor {Heterostructures} and {Nanostructures}},
    title = {Quantum {Transport} in {Semiconductor} {Nanostructures}},
    volume = {44},
    url = {https://www.sciencedirect.com/science/article/pii/S0081194708600910},
    urldate = {2025-10-31},
    booktitle = {Solid {State} {Physics}},
    publisher = {Academic Press},
    author = {Beenakker, C. W. J. and van Houten, H.},
    editor = {Ehrenreich, Henry and Turnbull, David},
    month = jan,
    year = {1991},
    doi = {10.1016/S0081-1947(08)60091-0},
    pages = {1--228},
}

@article{casanova_deep_2010,
    title = {Deep {Strong} {Coupling} {Regime} of the {Jaynes}-{Cummings} {Model}},
    volume = {105},
    doi = {10.1103/PhysRevLett.105.263603},
    number = {26},
    journal = {Physical Review Letters},
    author = {Casanova, J.},
    year = {2010},
}

@article{lu_cavity_2025,
    title = {Cavity engineering of solid-state materials without external driving},
    volume = {17},
    copyright = {© 2025 Optica Publishing Group},
    issn = {1943-8206},
    url = {https://opg.optica.org/aop/abstract.cfm?uri=aop-17-2-441},
    doi = {10.1364/AOP.544138},
    language = {english},
    number = {2},
    urldate = {2025-11-03},
    journal = {Advances in Optics and Photonics},
    author = {Lu, I.-Te and Shin, Dongbin and Svendsen, Mark Kamper and Latini, Simone and Hübener, Hannes and Ruggenthaler, Michael and Rubio, Angel},
    month = jun,
    year = {2025},
    note = {Publisher: Optica Publishing Group},
    pages = {441--525}
}

@misc{kockum_ultrastrong_nodate,
    title = {Ultrastrong coupling between light and matter {\textbar} {Nature} {Reviews} {Physics}},
    url = {https://www.nature.com/articles/s42254-018-0006-2},
    urldate = {2022-11-14},
    author = {Kockum, Anton Frisk},
}

\end{document}